\documentclass{article}
\usepackage{amsfonts,latexsym}
\newcommand{\diff}{\mbox{Diff}_+(S^1)}
\newtheorem{theorem}{Theorem}
\newtheorem{lemma}{Lemma}
\title{Lagrangian time-discretization of the Hunter-Saxton equation}
\author{Alexei V. Penskoi\thanks{Centre de recherches math\'ematiques, 
Universit\'e de Montr\'eal,
C.~P.~6128, succ. Centre-ville, Montr\'eal, 
Qu\'ebec, H3C 3J7, Canada {\tt e-mail: penskoi@crm.umontreal.ca}}}
\date{}
\begin{document}
\maketitle

\abstract{We study Lagrangian time-discretizations of the Hunter-Saxton 
equation. Using the Moser-Veselov approach, we obtain such
discretizations defined on the Virasoro group and on the group of
orientation-preserving diffeomorphisms of the circle. We conjecture that
one of these discretizations is integrable.}

\noindent 2000 Mathematical Subject Classification 34K99, 22E65, 70H99


\section*{Introduction}
In 1991 J.~K.~Hunter and R.~Saxton considered~\cite{HS} the equation
\begin{equation}\label{hs}
(u_t+uu_x)_x=\frac{1}{2}u^2_x,
\end{equation}
which describes the propagation of weakly nonlinear orientation waves in 
a massive nematic liquid crystal director field.
A derivative with respect to $x$ of the Hunter-Saxton equation 
\begin{equation}\label{hsd}
(u_t+uu_x)_{xx}=(\frac{1}{2}u^2_x)_x,\quad\mbox{or simply}\quad 
u_{txx}=-2u_xu_{xx}-uu_{xxx},
\end{equation}
is also often called the Hunter-Saxton equation.
It is equation~(\ref{hsd}) that will
be central to the present paper.

Let us recall some results concerning the Hunter-Saxton equation.
Equation~(\ref{hs}) was solved in~\cite{HS} using the method of 
characteristics. A generalization of equation~(\ref{hs}) 
was studied by M.~V.~Pavlov~\cite{Pa} and it was also solved. 
Equation~(\ref{hsd}) was investigated  
by J.~Hunter and Y.~Zheng~\cite{HZ}, and it was proven 
that~(\ref{hsd}) is a completely integrable, bi-variational,
bi-Hamiltonian system.

The Hunter-Saxton equation is related to the
Korteweg-de~Vries and Camassa-Holm equations in many ways.
P.~Olver and P.~Rosenau described~\cite{OR} this relation using
tri-Hamiltonian structures and scaling arguments. 
R.~Beals, D.~H.~Sattinger and J.~Szmigielski described~\cite{BSS1,BSS2}
the scattering theory for all three equations in a unified way.
B.~Khesin and G.~Misio\l{}ek interpreted~\cite{KM} these three
equations as the Euler equations describing
geodesic flows associated to different right-invariant metrics 
on the Virasoro group, or on an appropriate homogeneous space. Moreover, they
proved that these three equations exhaust all generic bi-Hamiltonian
systems related to the Virasoro group. The paper~\cite{KM} was a motivation for
the present paper and we will recall some results of~\cite{KM} in 
Section~\ref{hsandv}.

Since the Hunter-Saxton equation~(\ref{hsd}) is an Euler equation, it is 
natural to try to discretize it using the Moser-Veselov approach. This
approach is the following.

Let $M$ be a manifold, let $L$ be a function on $M\times M.$
A discrete Lagrangian system (see the review~\cite{V3} for references) 
describes stationary points of a functional
$S=S(X)$ defined on the space of sequences 
$X=(x_k),x_k\in M,k\in{\mathbb Z},$ by
a formal sum
$$
S(X)=\sum\limits_{k\in{\mathbb Z}}L(x_k,x_{k+1}).
$$
The function $L$ is called the Lagrangian. 

Let us assume that $M$ is a Lie group $G.$ It was observed by A.~P.~Veselov
and J.~Moser that symmetric
$$
L(x,y)=L(y,x)
$$
and right-invariant (or left-invariant)
$$
\forall g\in G\quad L(xg,yg)=L(x,y)
$$
Lagrangians often correspond to integrable systems which are
discretizations of the Euler equations corresponding to some right-invariant
(or left-invariant) metrics on $G.$ The case of $SO(n)$ and other
classical finite-dimensional groups $U(n),$ $Sp(n)$  
was studied extensively in~\cite{V1,V2,MV1}.
The first example of an infinite-dimensional group, namely
$\mbox{SDiff}({\mathbb R}^2)$
was considered in~\cite{MV2,FV}.

We shall study discrete Lagrangian systems on the Virasoro group
in order to find discretizations of the Hunter-Saxton equation. 
Discrete Lagrangian systems on the Virasoro group 
were studied in~\cite{P1,P2}. These papers were motivated by the observation
by V.~Yu.~Ovsienko and B.~A.~Khesin~\cite{KO} that the Korteweg-de~Vries
equation is the Euler equation for a right-invariant metric on the
Virasoro group. In~\cite{P1,P2} we investigated 
discrete Lagrangian systems on the Virasoro group trying to find
a time-discretization of the Korteweg-de~Vries equation. In~\cite{P1} we 
studied a particular class of discrete Lagrangian systems containing
an interesting system with remarkable properties. It was noticed 
however that this system had no visible relation to the Korteweg-de~Vries 
equation. We will show in the present paper that this system is in fact 
related to the Hunter-Saxton equation.
In~\cite{P2} we considered another particular class 
of discrete Lagrangian systems which has the Korteweg-de~Vries equation 
as a continuous limit. One of these systems has very interesting properties
and we conjectured that this system 
is integrable, but in fact till now only one integral was found.

In this paper we study a class of 
discrete Lagrangian systems on the Virasoro group
satisfying natural conditions generalizing those considered in~\cite{P1}. 
We show
that these systems have the Hunter-Saxton equation as their continuous limit.
We show that the system considered in~\cite{P1} which belongs to this 
class can be considered as a good candidate for a natural integrable
Lagrangian time-discretization of 
the Hunter-Saxton equation.

It is shown in~\cite{KM} that
the Hunter-Saxton equation can also be considered as an Euler equation
on the homogeneous space $\mbox{Diff}_{+}(S^1)/\mbox{Rot}(S^1)$ 
of the group of orientation-preserving diffeomorphisms 
of the circle modulo the group of rotations of the circle. We will also
consider possible discretizations of the Hunter-Saxton equation on the group 
$\diff.$ However, our primary 
interest will be the Virasoro group since it plays
a more fundamental role for the Hunter-Saxton equation. Moreover, we will see
that in some sense discrete Lagrangian systems on $\diff$ are
``reductions'' of discrete Lagrangian systems on the Virasoro group.

The plan of the paper is as follows. In Section~\ref{virga} we recall 
necessary facts about the Virasoro group and the Virasoro algebra.
In Section~\ref{hsandv} we recall the results of paper~\cite{KM}
describing links between the Hunter-Saxton equation and the
Virasoro group. In Section~\ref{discr} we study discrete Lagrangian 
systems on the Virasoro group. In Section~\ref{natural} we recall
the results of~\cite{P1} and show that the system studied in~\cite{P1}
can be considered as a good candidate for a natural integrable Lagrangian
time-discretization of the Hunter-Saxton equation. In the final 
Section~\ref{diff} we will study discrete Lagrangian 
systems on the group $\mbox{Diff}_{+}(S^1)$ and their relation to the
Hunter-Saxton equation. We will see that the ``reduction'' to $\diff$
of the system studied in Section~\ref{natural} is particularly simple.

\section{The Virasoro group and the Virasoro algebra}\label{virga}

Let  $\diff$ be the group of diffeomorphisms of $S^1$ preserving 
the orientation. We shall represent an element of $\diff$ as a diffeomorphism
$f:{\mathbb R}\rightarrow{\mathbb R}$ such that 
\begin{enumerate}
\item $f\in C^\infty({\mathbb R}),$
\item $f'(x)>0,$
\item $f(x+2\pi)=f(x)+2\pi.$
\end{enumerate}
Such a representation is not unique. Indeed, the functions 
$f+2\pi k,k\in\mathbb{Z}$ 
represent one element of $\diff.$

There exists a non-trivial central extension of
$\diff$ which is unique up to an isomorphism

This extension is called the Virasoro group (or the Bott-Virasoro group) 
and is denoted by
$\mbox{Vir}.$ Elements of $\mbox{Vir}$ are pairs $(f,F)$, 
where $f\in\diff,$ $F\in{\mathbb R}.$
The product of two elements is defined 
with the help of the Bott cocycle
as
$$
(f,F)\circ(g,G)=(f\circ g,F+G+\int\limits_{0}^{2\pi}\log(f\circ g)'\,d\log g').
$$
The unit element of $\mbox{Vir}$ is $(id,0).$ The inverse element of $(f,F)$ is
$(f^{-1},-F).$

The Virasoro algebra $\mbox{vir}$
is a Lie algebra corresponding to the Virasoro group. It is the 
central extension of the algebra $vect(S^1)$ of vector fields on 
the circle $S^1$
$$
\mbox{vir}=vect(S^1)\oplus{\mathbb R}.
$$
We represent an element of $vect(S^1)$ as $v(x)\partial_x,$ where
$v$ is a $2\pi$-periodic function. Thus an element of the Virasoro
algebra is a pair $(v(x)\partial_x,a).$ The algebra commutator in $\mbox{vir}$
is defined 
with the help of the Gelfand-Fuchs cocycle as
$$
[(v(x)\partial_x,a],[w(x)\partial_x,b])=%
((-vw_x+v_xw)(x)\partial_x,\int\limits_0^{2\pi}v_{xxx}w\,dx).
$$

\section{The Hunter-Saxton equation and the Virasoro group}\label{hsandv}

In this section we briefly recall some results of~\cite{KM} concerning
a link between the Hunter-Saxton equation and the Virasoro group.

Let $G$ be a Lie group, $\mathfrak{g}$ its Lie algebra. A right-invariant
metric on $G$ is completely defined by its restriction to $\mathfrak{g}$
$$
E(v)=\frac{1}{2}\langle v,Av\rangle,\quad v\in\mathfrak{g},
$$
where $A$ is a linear map $A:\mathfrak{g}\longrightarrow\mathfrak{g}^*$ called
the inertia operator.

To describe a geodesic $g(t)$ on $G,$ we transport its velocity vector to the 
identity by the right translation
$$
v(t)=R_{g^{-1}(t)*}\frac{d}{dt}g(t).
$$
Since $v(t)$ is an element of $\mathfrak{g},$ we can consider $m=Av$ which
is an element of the dual space $\mathfrak{g}^*.$ Then $m$ satisfies
the Euler equation given by the following explicit formula:
\begin{equation}\label{euler}
\frac{dm}{dt}=-ad^*_{A^{-1}m}m.
\end{equation}
This is a standard result which can be found in~\cite{A}. If we start with
a left-invariant metric, the sign in~(\ref{euler}) is reversed.

Let us now compute $ad^*$ for the Virasoro algebra. Let a space
$$
\{(u(dx)^2,a)|u\in C^\infty,u(x)=u(x+2\pi),a\in\mathbb{R}\}
$$
be the dual space $\mbox{vir}^*$ 
to the Virasoro algebra with the natural pairing
given by
$$
\langle(u(dx)^2,a),(w\partial_x,c)\rangle=\int\limits_0^{2\pi}uw\,dx+ac.
$$
Thus, $ad^*$ is defined by the formula
$$
\langle ad^*_{(v\partial_x,b)}(u(dx)^2,a),(w\partial_x,c)\rangle=%
\langle(u(dx)^2,a),[(v\partial_x,b),(w\partial_x,c)]\rangle.
$$
Using the definition of the commutator in $\mbox{vir}$ 
and integration by parts we 
obtain that the right-hand-side is equal to
$$
\int\limits_0^{2\pi}w(2uv_x+u_xv-av_{xxx})\,dx.
$$
Thus we obtain
$$
ad^*_{(v\partial_x,b)}(u(dx)^2,a)=((2uv_x+u_xv-av_{xxx})(dx)^2,0).
$$
Now let us consider an inertia operator 
$A:\mbox{vir}\longrightarrow\mbox{vir}^*,$ given
by
$$
A(v\partial_x,b)=((-\Lambda v)(dx)^2,b),
$$
where $\Lambda=-\partial_x^2.$ 
The corresponding scalar product on $\mathfrak{g}$ is given by the formula
$$
E((v\partial_x,b))=-\int\limits_0^{2\pi}vv_{xx}\,dx+b^2.
$$
Using integration by parts, we obtain
\begin{equation}\label{energy}
E((v\partial_x,b))=\int\limits_0^{2\pi}(v_x)^2\,dx+b^2.
\end{equation}
In the Euler equation~(\ref{euler})
we have $A^{-1}m$, but
our inertia operator $A$ is degenerate on $\mbox{vir}$
since $\Lambda$ has a non-trivial kernel consisting of constant vector fields 
on $S^1.$ It is necessary to consider the quotient space 
$\mbox{Vir}/Rot(S^1)$ of the 
Virasoro group by the subgroup of rotations of the circle. This is not
difficult since it is sufficient to consider $m$ only from the image
of the inertia operator, see~\cite{KM} for details.

We obtain the Euler equation~(\ref{euler}) on $\mbox{vir}^*$
$$
\frac{d}{dt}(u(dx)^2,a)=-ad^*_{A^{-1}(u(dx)^2,a)}(u(dx)^2,a)=
$$
$$
=-((2u\Lambda^{-1}u_x+u_x\Lambda^{-1}u-a\Lambda^{-1}u_{xxx})(dx)^2,0).
$$
If we put $v=\Lambda^{-1}u$ we obtain two equations: 
$$a_t=0$$  
(so $a$ is a constant) and
\begin{equation}\label{hsfromvir}
v_{txx}=-2v_xv_{xx}-vv_{xxx}-av_{xxx}.
\end{equation}
After the change of variables $u(x,t)=v(x,t)+a$ we obtain the Hunter-Saxton
equation~(\ref{hsd}).

It has also been shown~\cite{KM} that the Hunter-Saxton equation
can be considered as an Euler equation
on the homogeneous space $\mbox{Diff}_{+}(S^1)/\mbox{Rot}(S^1)$ 
of the group of orientation-preserving diffeomorphisms 
of the circle modulo the group of rotations of the circle.
The construction is quite analogous to the construction on the Virasoro group.
It is necessary to consider an inertia operator 
$A:vect(S^1)\longrightarrow vect(S^1)^*$ given by
$$
A(v\partial_x)=(-\Lambda v)(dx)^2.
$$
This inertia operator will give an equation
$$
v_{txx}=-2v_xv_{xx}-vv_{xxx},
$$
i.e. the Hunter-Saxton equation.
This coincides with equation~(\ref{hsfromvir}) for $a=0.$
We will see that in the discrete case the situation is analogous.

\section{Discrete Lagrangian systems on the Virasoro group}\label{discr}

Let $M$ be a manifold, let $L$ be a function on $M\times M.$
A discrete Lagrangian system describes stationary points of a functional
$S=S(X)$ defined on the space of sequences 
$X=(x_k),x_k\in M,k\in{\mathbb Z},$ by
a formal sum
$$
S(X)=\sum\limits_{k\in{\mathbb Z}}L(x_k,x_{k+1}).
$$
The function $L$ is called the Lagrangian.

Let us assume that $M$ is a Lie group $G.$
Our goal is to investigate potential candidates for a time-discretization
of the Euler equation. Since the Euler equation describes geodesics on
a Lie group equipped with right-invariant metric, it is natural to 
require that the function $L$ is right-invariant
$$
\forall g\in G\quad L(xg,yg)=L(x,y).
$$
Let $H(x)=L(x,e),$ where $e$ is the identity element of $G.$
A right-symmetric Lagrangian $L(x,y)$ is completely determined by $H$.
Indeed, $L(x,y)=L(xy^{-1},e)=H(xy^{-1}).$

We shall consider a class of functions of the type
\begin{equation}\label{functionH}
H((f,F))=F^2+\int\limits_0^{2\pi}U(f'(x))\,dx,
\end{equation}
where $f$ is a diffeomorphism, $F\in\mathbb{R}.$ 
It is a quite natural class since the scalar product in $\mbox{vir}$ defining
the Hunter-Saxton equation~(\ref{energy}) depends only on a derivative
of a vector field.

Let us consider a functional
$$
S=\sum\limits_{k\in{\mathbb Z}}L((f_k,F_k),(f_{k+1},F_{k+1})),
$$
where $\{(f_k,F_k)\}$ is a sequence of points on $\mbox{Vir}$ 
and $L$ is defined
using the function $H$~(\ref{functionH}) as described above:
$$
L((f_k,F_k),(f_{k+1},F_{k+1}))=H((f_k,F_k)\circ(f_{k+1},F_{k+1})^{-1}).
$$

\begin{theorem} 
The discrete Euler-Lagrange equations 
$\frac{\delta S}{\delta (f_k,F_k)}=0$ are the following:
\begin{equation}\label{EL1}
-\Omega_k+\Omega_{k+1}=0,
\end{equation}
$$
-2\Omega_k\left(\log(\omega_k')\right)''+U''(\omega_k')\omega_k'\omega_k''+
$$
\begin{equation}\label{EL2}
+2\Omega_{k+1}\left(\log((\omega_{k+1}^{-1})')\right)''+%
U''\left(\frac{1}{(\omega_{k+1}^{-1})'}\right)%
\frac{(\omega_{k+1}^{-1})''}{((\omega_{k+1}^{-1})')^2}=0,
\end{equation}
where $(\omega_k,\Omega_k)$ and $(\omega_{k+1},\Omega_{k+1})$  
are discrete analogues of angular velocities,
$$
(\omega_l,\Omega_l)=(f_{l-1},F_{l-1})\circ(f_l,F_l)^{-1},\quad l\in\mathbb{Z}.
$$
\end{theorem}

\noindent{\bf Proof.} It sufficient to consider a variation of the form
$$
(id+\varepsilon v,\varepsilon A)\circ(f_k,F_k)
$$
and to find a Taylor series of the variation of
$$
L((f_{k-1},F_{k-1}),(f_k,F_k))+L((f_k,F_k),(f_{k+1},F_{k+1}))
$$
up to order $O(\varepsilon^2).$ 
We obtain
$$
L((f_{k-1},F_{k-1}),(id+\varepsilon v,\varepsilon A)\circ(f_k,F_k))+
$$
$$
+L((id+\varepsilon v,\varepsilon A)\circ(f_k,F_k),(f_{k+1},F_{k+1}))=
$$
$$
=L((f_{k-1},F_{k-1}),(f_k,F_k))+L((f_k,F_k),(f_{k+1},F_{k+1}))+
$$
$$
+\varepsilon A(-2\Omega_k+2\Omega_{k+1})+
+\varepsilon\int\limits_0^{2\pi}%
\left[-2\Omega_k\left(\log(\omega_k')\right)''+%
U''(\omega_k')\omega_k'\omega_k''+\vphantom{\frac{1}{(\omega_{k+1}^{-1})'}}%
\right.
$$
$$
\left.+2\Omega_{k+1}\left(\log((\omega_{k+1}^{-1})')\right)''+%
U''\left(\frac{1}{(\omega_{k+1}^{-1})'}\right)%
\frac{(\omega_{k+1}^{-1})''}{((\omega_{k+1}^{-1})')^2}\right]v\,dx+%
O(\varepsilon^2)=0.
$$
Since $A$ is an arbitrary constant and $v$ is an 
arbitrary periodic function, we
obtain our formulae for the discrete Euler-Lagrange 
equation~(\ref{EL1},\ref{EL2}).
$\Box$

The first equation~(\ref{EL1}) is quite simple: it says that
$\Omega_{k+1}=\Omega_k,$ so $\Omega_k$ is an integral of our discrete 
Lagrangian system. The second equation is quite complicated.
It is necessary to remark that this is not a differential-difference equation.
Indeed, it includes $\omega_{k+1}^{-1}$ which is a diffeomorphism inverse to
$\omega_{k+1}.$ So, it is a complicated relation between $\omega_k$ and
$\omega_{k+1},$ and it is better to view it as a correspondence 
(i.~e. multivalued mapping) 
$$
\omega_k\mapsto \omega_{k+1}
$$
from $\diff$ to $\diff.$ 

Let us now find a continuous limit of our equations~(\ref{EL1},\ref{EL2}).
Our definition of a continuous limit is the following. Firstly, we suppose
that the angular velocity is of the form
$$
(\omega_l,\Omega_l)=(id+\varepsilon v_l(x),\varepsilon A_l), 
$$
i.~e. the angular velocity is the identity element of 
$\mbox{Vir}$ up to $O(\varepsilon).$
Secondly, we suppose that
\begin{equation}\label{cl}
v_k(x)=v(x,t),\quad A_k=A(t),\quad v_{k+1}(x)=v(x,t+\varepsilon),\quad%
A_{k+1}=A(t+\varepsilon).
\end{equation}
This is a quite natural discretization of time.
We substitute these formulae in the Euler-Lagrange 
equations~(\ref{EL1},\ref{EL2}), find a Taylor series, and finally
we define a continuous limit as an equation arising in this Taylor
series as the term of lowest order with respect to $\varepsilon.$

\begin{theorem} Let $U''(1)\ne 0.$
Then the continuous limit of the Euler-Lagrange 
equations~(\ref{EL1},\ref{EL2}) is the following:
\begin{equation}\label{cl1}
\Omega=const,
\end{equation}
\begin{equation}\label{cl2}
v_{txx}=2v_xv_{xx}+vv_{xxx}-\frac{4\Omega}{U''(1)}v_{xxx}.
\end{equation}
\end{theorem}

\noindent{\bf Proof.} We substitute the formulae~(\ref{cl})
in the equations~(\ref{EL1},\ref{EL2}) and we use a Taylor series.
From the equation~(\ref{EL1}) we obtain
$$
\varepsilon^2\frac{d}{dt}A(t)+O(\varepsilon^3)=0,
$$
thus $A(t)$ is a constant. Let us denote it by $\Omega.$ This gives us 
equation~(\ref{cl1}).
From equation~(\ref{EL2}) we obtain
$$
\varepsilon^2(-4\Omega v_{xxx}(x,t)+2U''(1)v_x(x,t)v_{xx}(x,t)+%
U''(1)vv_{xxx}-
$$
$$
-U''(1)v_{txx}(x,t))+O(\varepsilon^3)=0.
$$
Since $U''(1)\ne 0,$ this gives us equation~(\ref{cl2}).
$\Box$

If we do a simple change of time $t\mapsto -t$ and change our constant
$a=-\frac{4\Omega}{U''(1)},$ we obtain the Hunter-Saxton equation exactly in 
the same form as this equation arises as the Euler equation~(\ref{hsfromvir}):
$$
v_{txx}=-2v_xv_{xx}-vv_{xxx}-av_{xxx}.
$$

As we see, a wide class of discrete Lagrangian systems has the Hunter-Saxton
equation as a continuous limit. We have a natural question: how to
find a ``correct'' discretization? We will discuss it in the
next section.

\section{A candidate for a natural integrable Lagrangian 
time-discretization of the Hunter-Saxton equation}%
\label{natural}

Since the Hunter-Saxton equation is integrable, it is natural to 
suppose that a ``correct'' discretization is integrable in some sense.

To find a function $U$ giving an integrable discretization
we can use the observation by A.~P.~Veselov and 
J.~Moser~\cite{MV1,MV2} mentioned in the Introduction. Namely that 
known integrable Lagrangian
discretizations correspond to right-invariant (or left-invariant) 
{\em symmetric} Lagrangians:
$$
L(x,y)=L(y,x).
$$
Thus, let us also assume that our Lagrangian is symmetric.
The Lagrangian is given by the function $H:$
$$
L(x,y)=L(xy^{-1},e)=H(xy^{-1}).
$$
It is easy to see that symmetric Lagrangians correspond to inverse-invariant
function $H:$
$$
H(x^{-1})=H(x).
$$
Let us consider $H$ of the form~(\ref{functionH}). Which functions $U$ give us
an inverse-invariant $H$? This question was studied in~\cite{P1} and 
a sufficient condition was found:

\begin{lemma}~{\rm \cite{P1}} If a function $U$ satisfy the condition
\begin{equation}\label{condU}
xU\left(\frac{1}{x}\right)=U(x),
\end{equation}
then a function $H:\mbox{\rm Vir}\longrightarrow\mathbb{R}$ defined by  
formula~(\ref{functionH}) is inverse-invariant:
$$
H((f,F)^{-1})=H((f,F)).
$$
\end{lemma}

A second derivative of the identity~(\ref{condU}) is
$$
U''\left(\frac{1}{x}\right)\frac{1}{x^3}=U''(x).
$$
Using this identity we can rewrite equation~(\ref{EL2}) as
$$
\left[-2\Omega_k(\log((\omega_k)'))'-U'\left(\frac{1}{(\omega_k)'}\right)+%
\right.
$$
\begin{equation}\label{EL2A}
\left.+2\Omega_{k+1}(\log((\omega^{-1}_{k+1})'))'-%
U'\left(\frac{1}{(\omega^{-1}_{k+1})'}\right)\right]'=0.
\end{equation}
This is very symmetric expression 
and it is a complete derivative with respect to $x.$
We note that the Hunter-Saxton equation
arising from the Euler equation~(\ref{hsfromvir}) is also a complete 
derivative with respect to $x.$

The condition~(\ref{condU}) is very restrictive, however there are
many functions satisfying~(\ref{condU}). If we consider other cases studied,
there are no indications how to find an appropriate $U.$ 
In the case $G=SO(n)$~\cite{V1,V2,MV1} the final form of a Lagrangian
is obtained by imposing the condition that the Lagrangian $L(x,y)$ 
should be  bilinear in $x,$ $y.$ It is not 
clear what is the analog of this condition in our
case. In the case $G=\mbox{SDiff}({\mathbb R}^2)$~\cite{MV2,FV}
the Lagrangian is chosen as a natural generalization of 
the $SL(2)=Sp(2)$ case. It leads to an integrable system,
but there is no evidence that it is the unique Lagrangian with this property.

All we can do is to guess which function $U$ satisfying the 
condition~(\ref{condU}) gives us an integrable system. 
The answer to this question is still unknown, but we have a good candidate.
If we look at the equation~(\ref{EL2A}), we can see that if we have 
$\omega_k,$ we must solve a differential equation to find $\omega_{k+1}^{-1}$
and then use the inversion operation in the Virasoro group to find
$\omega_{k+1}.$ It is natural to suppose that an integrable system
corresponds to an integrable differential equation for
$\omega_{k+1}^{-1}.$ At least, analogous situation was observed 
in the case of $\mbox{SDiff}(\mathbb{R}^2)$~\cite{MV2}.
In~\cite{P1} 
we showed that a function $U(x)=\sqrt{x}$ gives us a system 
such that if we have $\omega_k$, we can find
$\omega_{k+1}$ by solving a linear first-order differential equation
and using the inversion operation in $\mbox{Vir}.$ In the rest of this 
section we follow~\cite{P1}.

Let us consider the equation~(\ref{EL2A}) with $U(x)=\sqrt{x}:$
$$
\left[-2\Omega_k(\log((\omega_k)'))'-%
\frac{1}{2}\sqrt{(\omega_k)'}+\right.
$$
\begin{equation}\label{E2B}
\left.+2\Omega_{k+1}(\log((\omega^{-1}_{k+1})'))'-%
\frac{1}{2}\sqrt{(\omega^{-1}_{k+1})'}\right]'=0.
\end{equation}
Let us integrate this equation once and put
$$
\Phi=\frac{1}{\sqrt{(\omega_k)'}},\quad%
\Psi=\frac{1}{\sqrt{(\omega^{-1}_{k+1})'}}.
$$
We obtain the equation
$$
8\Omega\left(-\frac{\Phi'}{\Phi}+\frac{\Psi'}{\Psi}\right)+%
\frac{1}{\Phi}+\frac{1}{\Psi}+C=0,
$$
where $C$ is a constant of integration and $\Omega=\Omega_k=\Omega_{k+1}$
(remember that $\Omega_k$ is an integral).
This equation is equivalent to the equation
$$
\Psi'+\Psi\left(\frac{C}{8\Omega}+\frac{1}{8\Omega\Phi}-%
\frac{\Phi'}{\Phi}\right)+\frac{1}{8\Omega}=0.
$$
This is a linear first-order differential equation for $\Psi$ with
periodic coefficients depending on $\Phi.$ For generic $\Phi$
it has only one solution, so $\Psi$ is determined by $\Phi$ up to
a constant $C.$ Reconstructing $\omega^{-1}_{k+1}$ 
from $\Psi$ we obtain another
constant, so we have a following result: $\omega_{k+1}$ is obtained
from $\omega_{k}$ by a two-parametric correspondence. To
find $\omega_{k+1}$ starting from $\omega_k$, we must solve a first-order
linear differential equation to find $\omega^{-1}_{k+1},$ 
and then reconstruct $\omega_{k+1}$ from
$\omega^{-1}_{k+1}$ by inversion. 
As mentioned before, the analogous situation was observed in the case
of $\mbox{SDiff}({\mathbb R}^2)$~\cite{MV2} which is integrable~\cite{FV}.
For this reason we consider system~(\ref{E2B})
to be a good candidate for an integrable Lagrangian discretization of the
Hunter-Saxton equation.

\section{Discrete Lagrangian systems on the group of 
orientation-preserving diffeomorphisms of the circle}\label{diff}

As explained in Section~\ref{hsandv}, we can obtain the Hunter-Saxton
equation using not only the Virasoro group, but also the group
$\mbox{Diff}_+(S^1),$ and the construction is quite 
analogous. Similarly, we can consider discrete Lagrangian systems on the 
group $\mbox{Diff}_+(S^1)$ and obtain results analogous to the results for 
the Virasoro group. In this Section we formulate these results, proofs are 
omitted since they are analogous to the proofs of the last two sections.

We shall consider a class of functions 
$H:\diff\longrightarrow\mathbb{R}$
of the type
\begin{equation}\label{functionHDiff}
H(f)=\int\limits_0^{2\pi}U(f'(x))\,dx,
\end{equation}
where $f$ is a diffeomorphism.

Let us consider a functional
$$
S=\sum\limits_{k\in{\mathbb Z}}L(f_k,f_{k+1}),
$$
where $\{f_k\}$ is a sequence of points on $\mbox{Diff}_+(S^1)$ 
and $L$ is defined
using the function $H$~(\ref{functionHDiff}):
$$
L(f_k,f_{k+1})=H(f_k\circ f_{k+1}^{-1}).
$$

\begin{theorem} 
The discrete Euler-Lagrange equation 
$\frac{\delta S}{\delta f_k}=0$ is the following:
\begin{equation}\label{ELDiff}
U''(\omega_k')\omega_k'\omega_k''+
U''\left(\frac{1}{(\omega_{k+1}^{-1})'}\right)%
\frac{(\omega_{k+1}^{-1})''}{((\omega_{k+1}^{-1})')^2}=0,
\end{equation}
where $\omega_k$ and $\omega_{k+1}$  
are discrete analogues of angular velocities,
$$
\omega_l=f_{l-1}\circ f_l^{-1},\quad l\in\mathbb{Z}.
$$
\end{theorem}

This is exactly equation~(\ref{EL2}) with $\Omega_k=\Omega_{k+1}=0.$
It is similar to the relation between the Euler equations on 
$\mbox{Vir}/Rot(S^1)$ and $\diff/Rot(S^1),$ see the end of 
Section~\ref{hsandv}.
Thus, Lagrangian discrete systems on $\diff$ are in this sense
the ``reductions'' of Lagrangian discrete systems on the Virasoro group.
Using this remark, it is easy to find a continuous limit of
equation~(\ref{ELDiff}).

\begin{theorem} Let $U''(1)\ne 0.$
Then the continuous limit of the Euler-Lagrange 
equation~(\ref{ELDiff}) is the following:
$$
v_{txx}=2v_xv_{xx}+vv_{xxx}.
$$
\end{theorem}

\begin{lemma} If a function $U$ satisfies the condition~(\ref{condU})
then a function $H:\diff\longrightarrow\mathbb{R}$ defined by  
formula~(\ref{functionHDiff}) is inverse-invariant:
$$
H(f^{-1})=H(f).
$$
\end{lemma}

If the function U satisfies the condition~(\ref{condU}), we can rewrite
equation~(\ref{ELDiff}) as
\begin{equation}\label{ELADiff}
\left[U'\left(\frac{1}{(\omega_k)'}\right)+%
U'\left(\frac{1}{(\omega^{-1}_{k+1})'}\right)\right]'=0.
\end{equation}

As in the previous Section let us consider $U(x)=\sqrt{x}.$ 
Equation~(\ref{ELADiff}) becomes particularly simple
\begin{equation}\label{zvv}
\left[\sqrt{(\omega_k)'}+%
\sqrt{(\omega^{-1}_{k+1})'}\right]'=0.
\end{equation}

To find $\omega_{k+1}$ starting from $\omega_k$, we can rewrite the
previous equation as
$$
(\omega^{-1}_{k+1})'=(C-\sqrt{(\omega_k)'})^2,
$$
where $C$ is an arbitrary constant.
We see that given $\omega_k,$
we can find $\omega_{k+1}$ using integration and the inversion 
operation in the group $\diff.$
We see that system~(\ref{zvv}) has the same property
as the the equation~(\ref{E2B}) and
can be also considered  as a good
candidate for an integrable Lagrangian discretization of the
Hunter-Saxton equation.

\section{Conclusions}

We studied discrete Lagrangian systems on the Virasoro group and
the group of orientation-preserving diffeomorphisms of the circle. 
It is shown that under some very natural assumptions, these systems
have the Hunter-Saxton equation as their continuous limit. In particular
we studied a special discrete Lagrangian system on the Virasoro group
which has the following properties:
\begin{enumerate}
\item Its continuous limit is the Hunter-Saxton equation.\label{p1}
\item It can be solved by solving a first-order linear differential 
equation and using the inversion operation in the Virasoro group.\label{p2}
\item The Lagrangian is symmetric and right-invariant.\label{p3} 
\end{enumerate} 
The ``reduction'' of this system to $\diff$ has properties~\ref{p1} 
and~\ref{p3}, but~\ref{p2} is replaced but simpler property: 
it can be solved by integration and 
the inversion operation in the group $\diff$.

In the case of $\mbox{SDiff}(\mathbb{R}^2)$ studied in~\cite{MV2,FV}
the Lagrangian discrete system is integrable (in the sense that
the dynamics are linearized). We conjecture that the same is true
for two systems studied in this paper since their properties
are analogous to the properties of the system on $\mbox{SDiff}(\mathbb{R}^2)$.
 
The Lagrangian discrete system on $\mbox{SDiff}(\mathbb{R}^2)$
can be interpreted as a chain of B\"acklund
transformations for the (integrable) Monge-Amp\`ere equation~\cite{MV2}. 
This is nowadays a standard way to think about integrable discretizations. 
It would be very interesting to prove similar theorems for the systems 
discussed in this paper.

\section*{Acknowledgments}

The author is very grateful to the Centre de Recherches Math\'ematiques (CRM)
for its hospitality. He thanks Professor A.~P.~Veselov,
who drew his attention to Lagrangian discrete systems many years
ago, for very useful discussions which greatly improved this paper.
The author is very grateful to Professor
B.~A.~Khesin for stimulating discussions during his short stay at the CRM 
and for giving him the preprint~\cite{KM}. The author would like to
thank Professor P.~Winternitz for useful discussions.
The author also would like to thank 
Professor J.~Szmigielski for sending the preprint~\cite{BSS2}.

\end{document}